\documentclass[
	aps,
	amsmath,amssymb,
	reprint,
	superscriptaddress,
	pra
   ]{revtex4-2}

\newcommand{\Tr}{\mathrm{Tr}}

\usepackage{enumerate}
\usepackage{paralist}
\usepackage{graphicx, color}
\usepackage{comment}
\usepackage{dcolumn}
\usepackage{siunitx}
\usepackage{bm}
\usepackage{braket}
\usepackage{here}

\usepackage{here}
\begin{document}

\title{Learning temporal data  with variational quantum recurrent neural network}

\author{Yuto Takaki}
\affiliation{Graduate School of Engineering Science, Osaka University, 1-3 Machikaneyama, Toyonaka, Osaka 560-8531, Japan}

\author{Kosuke Mitarai}
\email{mitarai@qc.ee.es.osaka-u.ac.jp}
\affiliation{Graduate School of Engineering Science, Osaka University, 1-3 Machikaneyama, Toyonaka, Osaka 560-8531, Japan}
\affiliation{Center for Quantum Information and Quantum Biology, Institute for Open and Transdisciplinary Research Initiatives, Osaka University, Japan}
\affiliation{JST, PRESTO, 4-1-8 Honcho, Kawaguchi, Saitama 332-0012, Japan}

\author{Makoto Negoro}
\affiliation{Center for Quantum Information and Quantum Biology, Institute for Open and Transdisciplinary Research Initiatives, Osaka University, Japan}
\affiliation{Institute for Quantum Life Science, National Institutes for Quantum and Radiological Science and Technology, Japan}

\author{Keisuke Fujii}
\affiliation{Graduate School of Engineering Science, Osaka University, 1-3 Machikaneyama, Toyonaka, Osaka 560-8531, Japan}
\affiliation{Center for Quantum Information and Quantum Biology, Institute for Open and Transdisciplinary Research Initiatives, Osaka University, Japan}
\affiliation{Center for Emergent Matter Science, RIKEN, Wako Saitama 351-0198, Japan}

\author{Masahiro Kitagawa}
\affiliation{Graduate School of Engineering Science, Osaka University, 1-3 Machikaneyama, Toyonaka, Osaka 560-8531, Japan}
\affiliation{Center for Quantum Information and Quantum Biology, Institute for Open and Transdisciplinary Research Initiatives, Osaka University, Japan}

\date{\today}

\begin{abstract}
	We propose a method for learning temporal data using a parametrized quantum circuit. 
We use the circuit that has a similar structure as the recurrent neural network which is one of the standard approaches employed for this type of machine learning task.
Some of the qubits in the circuit are utilized for memorizing past data, while others are measured and initialized at each time step for obtaining predictions and encoding a new input datum.
The proposed approach utilizes the tensor product structure to get nonlinearity with respect to the inputs.
Fully controllable, ensemble quantum systems such as an NMR quantum computer is a suitable choice of an experimental platform for this proposal. 
We demonstrate its capability with Simple numerical simulations, in which we test the proposed method for the task of predicting cosine and triangular waves and quantum spin dynamics.
Finally, we analyze the dependency of its performance on the interaction strength among the qubits in numerical simulation and find that there is an appropriate range of the strength.
This work provides a way to exploit complex quantum dynamics for learning temporal data.
\end{abstract}

\maketitle

\section{Introduction}
Quantum machine learning \cite{Biamonte2017} is gaining attention in the hope of speeding up machine learning tasks with quantum computers.
One direction is to develop quantum machine learning algorithms with proven speedups over classical approaches \cite{Schuld2016, Wiebe2012, Rebentrost2014, yamasaki2020learning}.
Another direction, which is becoming popular recently, is to construct heuristic algorithms using parametrized quantum circuits.
Among such, a popular idea is to use an exponentially large Hilbert space of a quantum system for storing features of dataset \cite{Schuld2019, Mitarai2018, farhi2018, cerezo2020variational}.
In this approach, each datum is first encoded in a quantum state, and then we employ a parametrized quantum circuit to extract features that are not classically tractable.

Here, we specifically consider the extension of the latter approaches for the learning of temporal data.
In practice, such machine learning tasks frequently appear in various fields such as natural language processing, speech recognition, and stock price prediction.
A previous idea of using a quantum system for the time-series analysis can be found in Ref. \cite{Fujii2017}, where the authors proposed quantum reservoir computing (QRC).
QRC employs complex quantum dynamics itself as a computational reservoir.
It learns to perform temporal tasks by optimizing the readout from the reservoir while the quantum system is left unchanged.
This is in contrast to the approach taken in this work, that is, we directly tune the dynamics of the quantum system in use.

In this work, we propose a type of parametrized quantum circuit for learning temporal data in analogy with the recurrent neural network (RNN) which is a popular machine learning model on a classical computer for the task \cite{SHERSTINSKY2020132306}.
We call the proposed circuit quantum recurrent neural network (QRNN).
In QRNN, we repeat the following three steps; encoding an input, applying a parametrized quantum circuit, and measuring a portion of qubits to obtain a prediction from the system.
The circuit applied at the second step remains the same at every loop, introducing a recurrent structure.
The parameters in the circuit are optimized with respect to a suitable cost function whose minimization leads to an accurate prediction of a given temporal dataset.
This algorithm employs the tensor product structure of quantum systems to obtain the nonlinearity of the output with respect to input in a way similar to Ref. \cite{Mitarai2018}.
To use expectation values to obtain the output prediction while maintaining the recurrent structure, we assume the use of an ensemble quantum system such as NMR quantum computers which allows us to efficiently extract them.
We also test the validity of the proposed method by numerically simulating the training and prediction processes.
We believe this work opens up a way to exploit parametrized quantum circuits for time-series analysis.

The rest of this work is organized as follows.
In Sec. \ref{sec:pre}, we first review the classical recurrent neural network and existing machine learning algorithms with parametrized quantum circuits.
Then, the proposed QRNN is described in detail in Sec. \ref{sec:QRNN}.
In Sec. \ref{sec:numerical}, we present numerical simulations where we apply QRNN to some simple tome-series prediction tasks and analyze results.
Conclusion and outlook are given in Sec. \ref{sec:conclusion}.

\section{Preliminary}\label{sec:pre}

\subsection{Learning of temporal data}
In a (supervised) temporal learning task, we are given with a input sequence $\{\bm{x}_t\}_{t=0}^{T-1}$ and a target teacher sequence $\{\bm{y}_t\}_{t=0}^{T-1}$ as a training dataset.
Our task is to construct a model $\overline{\bm{y}_k} = f(\{\bm{x}_t\}_{t=0}^k)$ such that difference between $\bm{y}_k$ and $f(\{\bm{x}_t\}_{t=0}^k)$ is small from the training dataset.
A popular example is the time-series prediction task where $\bm{y}_t=\bm{x}_{t+1}$, which will be used for demonstration of the proposed method in Sec. \ref{sec:numerical}.
After constructing a model that is accurate enough, we can perform the prediction of $\bm{y}_t$.

\subsection{Recurrent neural network}

\begin{figure}
    \centering
    \includegraphics[width=0.8\linewidth]{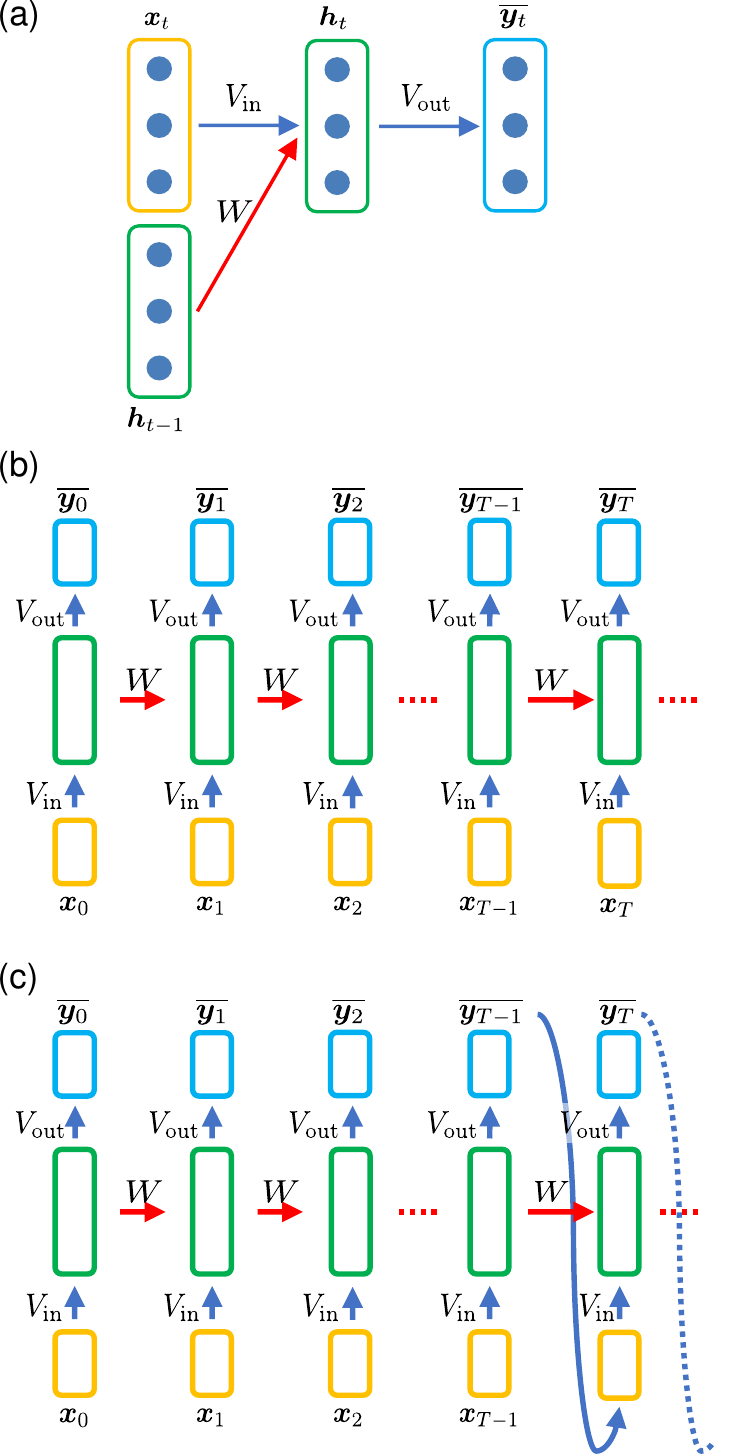}
    \caption{(a) Schematic picture of a basic RNN. Blue circles represent elements of vectors $\bm{x}_t$, $\bm{h}_t$ and $\overline{\bm{x}_t}$. The matrices $V_{\mathrm{in}}$, $V_{\mathrm{out}}$ and $W$ connecting layers are the parameters to be optimized during training of the model. The bias vectors $\bm{b}$ and $\bm{c}$ are abbreviated for simplicity. (b) Unfolded version of (a). (c) RNN for time-series prediction.}
    \label{fig:RNN}
\end{figure}

RNN \cite{SHERSTINSKY2020132306} is a famous technique to analyze time-series data.
There are a number of variants of RNN models such as Elman network, long short term memory (LSTM), and gated recurrent unit (GRU) \cite{hochreiter1997long, chung2014empirical, elman1990finding}.
Among those, we use a basic form of RNN shown in  Fig.~\ref{fig:RNN} (a) as a reference to construct its quantum version.

RNN consists of an input layer $\bm{x}_t$, hidden layer $\bm{h}_t$, and output layer $\overline{{\bm y}_{t+1}}$, which are real-valued vectors.
When $\bm{x}_t$ is injected into an input layer of RNN, the value of the hidden layer is updated using the value of $\bm{x}_t$ and $\bm{h}_{t-1}$ as,
\begin{align}
    \bm{h}_t = g_h(V_{\mathrm{in}}\bm{x}_t+W\bm{h}_{t-1}+\bm{b}_{\mathrm{in}})
\end{align}
where $V_{\mathrm{in}}$ and $W$ are matrices, $\bm{b}_{\mathrm{in}}$ is a bias vector, and $g_h(\cdot)$ is an activation function which element-wisely apply a nonlinear transformation to a vector.
The output $\overline{\bm{y}_{t+1}}$ is computed as,
\begin{align}
    \overline{\bm{y}_{t+1}} = g_o(V_{\mathrm{out}}\bm{h}_t+\bm{b}_{\mathrm{out}})
\end{align}
where $V_{\mathrm{out}}$, $\bm{b}_{\mathrm{out}}$, and $g_o(\cdot)$ are a matrix, a bias vector, and an activation function like $g_h$, respectively.
$W$, $V_{\mathrm{in}}$, $\bm{b}_{\mathrm{in}}$, $V_{\mathrm{out}}$, and $\bm{b}_{\mathrm{out}}$ are parameters to be optimized to minimize a cost function $L$ which represents the difference between $\bm{y}_{t+1}$ and $\overline{\bm{y}_{t+1}}$ such as the squared error $L=\sum_t \|\overline{\bm{y}_{t}}-\bm{y}_{t}\|^{2}$.
The recurrent structure where the value of the hidden layer at time $t$ is computed from $\bm{h}_{t-1}$ allows the RNN to hold the information given in the past time steps \cite{SHERSTINSKY2020132306}.

Frequently, the training of RNN is performed with the unfolded expression as shown in Fig. \ref{fig:RNN} (b), which allows us to treat RNN in the same manner as feed-forward neural networks and to use standard techniques such as backpropagation to optimize the parameters $W$, $V_{\mathrm{in}}$, $\bm{b}_{\mathrm{in}}$, $V_{\mathrm{out}}$, and $\bm{b}_{\mathrm{out}}$. 
After the training using the dataset $\{\bm{x}_t\}_{t=0}^{T-1}$ and $\{\bm{y}_t\}_{t=0}^{T-1}$, we can proceed to the prediction of $\bm{y}_T$, $\bm{y}_{T+1}$, $\cdots$ by repeating the structure.
However, for the time-series prediction task, the next input data $\bm{x}_T$ itself is unknown and to be predicted. 
Therefore, the prediction has to be performed using the strategy given in Fig.~\ref{fig:RNN} (c) where we feed its prediction  $\overline{\bm{y}_t}$ as the input.
This strategy is justified since we expect the trained network to output $\overline{\bm{y}_t}$ that is close to the true value $\bm{x}_{t+1}$. 
We follow the above structure closely to construct its quantum version in the following sections.

\subsection{Machine learning with parametrized quantum circuit}
Many researchers have developed a wide range of algorithms \cite{cerezo2020variational} using parametrized circuits, including quantum chemistry simulations \cite{Peruzzo2014}, combinational optimization problems \cite{farhi2014quantum}, and machine learning \cite{Mitarai2018, farhi2018, Schuld2019, Benedetti2019}.
Algorithms involving parametrized quantum circuits usually works in the following way.
First, we apply a parametrized quantum circuit $U(\bm{\theta})$ to some initialized state $\ket{\psi_0}$, where $\bm{\theta}$ is the parameter of the circuit.
Then, we measure an expectation value of a specific observable, $\braket{O(\bm{\theta})}$.
Based on the value, we compute a cost function $L(\braket{O(\bm{\theta})})$ to be minimized by optimizing $\bm{\theta}$.

For machine learning tasks, we define $L(\braket{O(\bm{\theta})})$ such that the output $\braket{O(\bm{\theta})}$ provides us appropriate predictions when it is minimized.
Let us describe an example of supervised learning \cite{Mitarai2018, farhi2018, Schuld2019}.
In supervised learning, we are provided with a training dataset consisting of input data $\{\bm{v}_i\}$ and corresponding teacher data $\{u_i\}$.
Ideas introduced in Refs. \cite{Mitarai2018, farhi2018, Schuld2019} are to use a parametrized quantum circuit $U(\bm{\theta},\bm{v}_i)$ which depends also on $\bm{v}_i$.
This circuit encodes an input $\bm{v}_i$ into a quantum state which results in the input-dependent output $\braket{O(\bm{\theta},\bm{v}_i)}$.
The cost function is defined as, for example, the mean squared error between $\{u_i\}$ and $\{\braket{O(\bm{\theta},\bm{v}_i)}\}$.
Minimization of such a cost function by optimizing the parameters $\bm{\theta}$ results in $\braket{O(\bm{\theta}_{\mathrm{opt}},\bm{v}_i)}$ that is close to $u_i$, where $\bm{\theta}_{\mathrm{opt}}$ is the optimized parameter.
Finally, we can use $\braket{O(\bm{\theta}_{\mathrm{opt}},\bm{v})}$ for an unknown input $\bm{v}$ to predict corresponding $u$.
In the following sections, we extend this approach to the task of predicting time-series data by combining it with the ideas of RNN.

\section{Quantum Recurrent Neural Network}\label{sec:QRNN}
In this section, we first describe the proposed algorithm and its theoretical capability.
Then, a possible experimental setup for its realization and its relation with previous works are discussed.

\subsection{Algorithm}
\label{sec:algorithm}

\begin{figure*}
    \centering
    \includegraphics[width=0.7\linewidth]{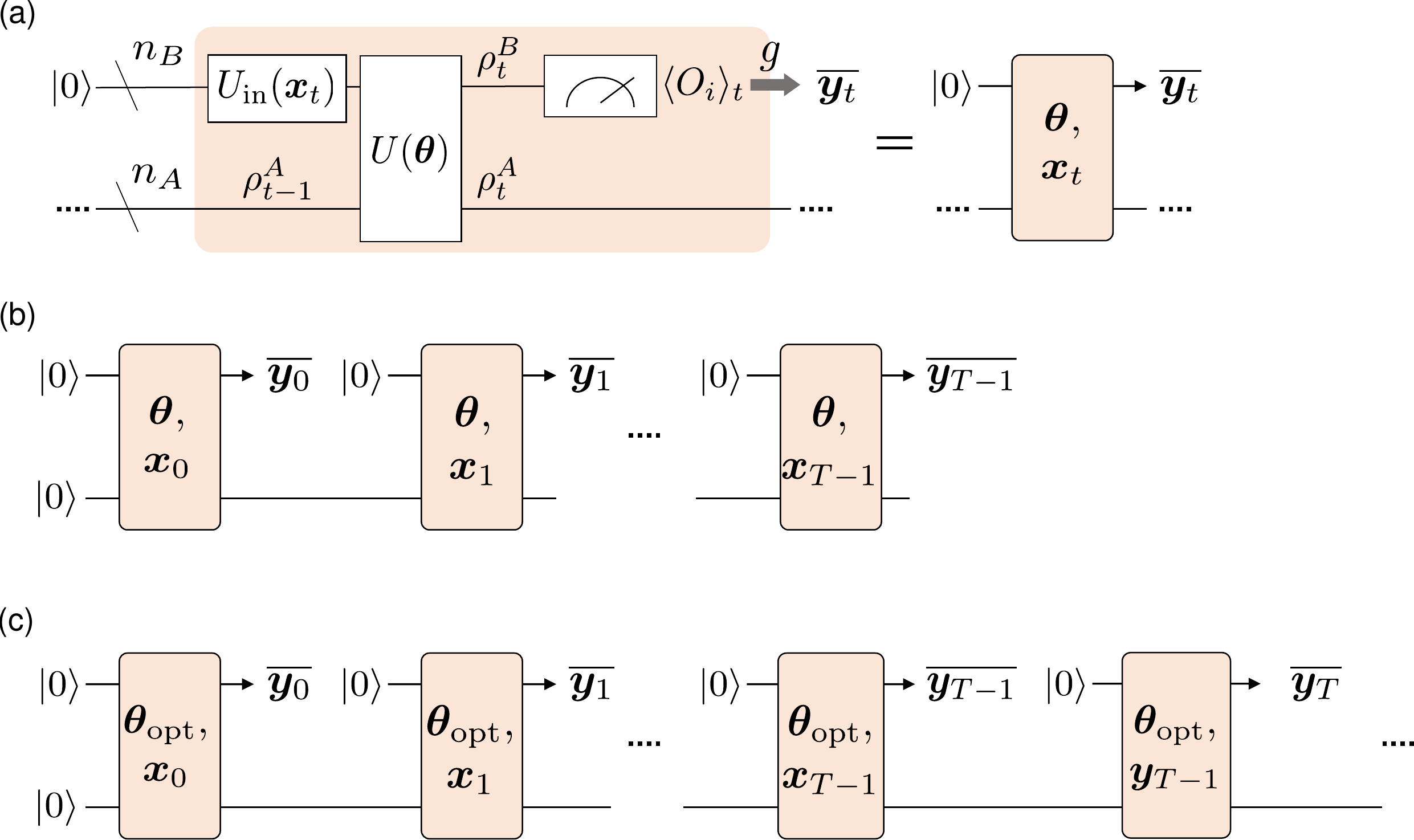}
    \caption{Structure of QRNN. (a) Structure of QRNN for a single time step.
    (b) QRNN in the training phase. 
    (c) QRNN in the prediction phase. Note that the obtained prediction $\overline{\bm{x}_T}$ is used as the input to the next time step.}
    \label{fig:qrnn}
\end{figure*}

Fig. \ref{fig:qrnn} shows a schematic of QRNN algorithm using $n=n_A+n_B$ qubits.
QRNN is composed of two groups of qubits called 'A' and 'B'.
Groups A and B have $n_A$ and $n_B$ qubits respectively.
The qubits in group A are never measured throughout the algorithm so that they hold past information.
On the other hand, group B qubits are measured and initialized at each time step $t$ to output the prediction and to take input to the system.

Each time step of QRNN consists of three parts, namely, encoding part, evolution part, and measurement part, which are schematically depicted in Fig.  \ref{fig:qrnn} (a).
At the encoding part of time step $t$, we encode a training datum $\bm{x}_{t}$ to the quantum state of qubits in group B by applying $U_{\rm in}(\bm{x}_{t})$ to the initilized state $\ket{0}^{\otimes n_B}$.
Hereafter, we abbreviate $\ket{0}^{\otimes n_B}$ as $\ket{0}$ if there is no confusion.
Note that group A already holds information about $\bm{x}_0, \cdots, \bm{x}_{t-1}$ as a density matrix $\rho^A_{t-1}(\bm{\theta}, \bm{x}_0, \cdots, \bm{x}_{t-1})$ resulted from previous steps.
At the evolution part, we apply a parameterized unitary circuit $U(\bm{\theta})$ to the entire qubits in the system.
$U(\bm{\theta})$ transfers the information injected in group B to group A by introducing the interaction among the qubits.
We denote the reduced density matrix of $A$ and $B$ as $\rho^A_t$ and $\rho^B_t$ after the evolution, respectively.
At the measurement part, we first measure expectation values of a set of commuting observables, $\{O_i\}$, of group B to obtain,
\begin{align}
    \braket{O_i}_{t}=\mathrm{Tr}[\rho^{B}_{t}O_i].
\end{align}
Then, we transform this expectation value to get the predicion $\overline{\bm{y}_{t}}$ of $\bm{y}_{t}$ by some function $g$;
$\overline{\bm{y}_{t}}=g\left(\{\braket{O_i}_{t}\}\right)$.
$g$ can, for example, be a linear combination of $\{\braket{O_i}_{t}\}$. 
Note that the transformation $g$ can be chosen arbitrarily and optimized in general.
Afterward, the qubits in group B are initialized to $\ket{0}$.

We repeat the three parts to obtain the predictions $\overline{\bm{y}_0}, \cdots, \overline{\bm{y}_{T-1}}$ for ${\bm{y}_0}, \cdots, {\bm{y}_{T-1}}$.
After obtaining the predictions, we compute the cost function $L(\{\bm{y}_0, \cdots, \bm{y}_{T-1}\}, \{\overline{\bm{y}_0}, \cdots, \overline{\bm{y}_{T-1}}\})$ which represents the difference between the training data $\{\bm{y}_0, \cdots, \bm{y}_{T-1}\}$ and prediction $\{\overline{\bm{y}_0}, \cdots, \overline{\bm{y}_{T-1}}\}$ obtained by QRNN.
The parameter $\bm{\theta}$ is optimized to minimize $L$.
This optimization is performed by standard optimizers running on a classical computer.
For example, we can utilize gradient-free optimization methods such as Nelder-Mead or simultaneous perturbation stochastic approximation (SPSA) algorithms.
Another choice would be to use a gradient-based method such as gradient descent.
The analytic gradient of the cost function can be obtained using the so-called parameter shift rule \cite{Mitarai2018, schuld2019evaluating, koczor2020quantum} by running QRNN for $O(T)$ times for each parameter, see Appendix for detail.

After the training, we expect the trained QRNN to be able to predict $\bm{y}_t$ with $\overline{\bm{y}}_t$ for $t\geq T$ after running it through $t=0$ to $T-1$.
For the time-series prediction task, we use its prediction $\overline{\bm{y}_t}$ as input to the system similarly to the ordinary RNN case, since we expect QRNN to output $\overline{\bm{y}_t}$ which is close to its true value $\bm{x}_{t+1}$.

\subsection{Capability of QRNN}

Here, we will discuss what types of function QRNN can model using a simple example.
QRNN exploits the tensor product structure to gain the nonlinearity of the output with respect to the input in a way similar to Ref. \cite{Mitarai2018}.

For simplicity, here we assume that the input and teacher data are scalar.
Suppose that we use a single qubit as group B, and take $U_{\mathrm{in}}(x) = R_y(\arccos x)$.
This gate applied to $\ket{0}$ encodes $x$ to the $Z$-expectation value of the qubit, that is, at $t=0$, the state after injecting the first input datum $x_0$ is,
\begin{equation}
\frac{1}{2}(I+\sqrt{1-x_0^2} X+ x_0 Z)\otimes \ket{0}\bra{0}^{\otimes n_A}.
\end{equation}
After the evolution $U(\bm{\theta})$, the density matrix of the system is,
\begin{equation}\label{eq:t0rho}
\sum_{P\in \mathcal{P}_n} \left(c_{1P}(\bm{\theta})x_0 + c_{2P}(\bm{\theta})\sqrt{1-x_0^2}+c_{3P}(\bm{\theta})\right) P,
\end{equation}
where $\mathcal{P}_n = \{I,X,Y,Z\}^{\otimes n}$ and $c_{1P}(\bm{\theta})$, $c_{2P}(\bm{\theta})$, and $c_{3P}(\bm{\theta})$ are real coefficients.
This means that an expectation value of any Pauli operator $P$ can be written as a linear combination of $\{x_0, \sqrt{x_0-1}, 1\}$, and the output $\overline{y_0}$ can also be written in terms of them.

In the next time step of QRNN, $\rho_{0}^A$, which is the reduced density matrix of Eq. (\ref{eq:t0rho}) obtained by tracing out $B$, is generally in the form of,
\begin{equation}\label{eq:rho_1A}
\rho_{0}^A = \sum_{P\in\mathcal{P}_{n_A}} \left(c_{1P}'(\bm{\theta})x_0 + c_{2P}'(\bm{\theta})\sqrt{1-x_0^2}+c_{3P}'(\bm{\theta})\right) P,
\end{equation}
with some coefficient $c_{1P}'(\bm{\theta})$, $c_{2P}'(\bm{\theta})$, and $c_{3P}'(\bm{\theta})$.
Then, $x_1$ is injected into $B$ by $U_{\mathrm{in}}(x)$, and $U(\bm{\theta})$ is applied to the whole system.
This results in the density matrix in the form of,
\begin{equation}\label{eq:t1rho}
\sum_{P\in\mathcal{P}_{n_A}} \sum_i c_{iP}''(\bm{\theta})\phi_i(x_0, x_1) P,
\end{equation}
where $\{\phi_i(x_0,x_1)\} = \{x_0,$ $x_1,$ $\sqrt{1-x_0^2},$ $\sqrt{1-x_1^2},$ $x_0x_1,$ $x_0\sqrt{1-x_1^2},$ $x_1\sqrt{1-x_0^2},$ $1\}$, and $c_{iP}''$ are coefficients.
Note that the nonlinear functions such as $x_0x_1$ originate from the tensor product structure of the system.
Therefore, we can conclude the output $\overline{y_1}$ can be written in terms of a linear combination of $\phi_i(x_0, x_1)$ in a way similar to the above.

Repeating the above discussion, we can see that the output of QRNN can have highly nonlinear terms such as $\prod_{t=0}^{\tau} x_t$ thanks to the tensor product structure.
Note that QRNN can ``choose'' which term to be kept in group A qubits by tuning $U(\bm{\theta})$.
For example, QRNN can fully keep $x_0$ term in $\rho_0^A$ if $U(\bm{\theta})$ transforms $X\otimes I^{\otimes n_A}$ to a local Pauli operator only acting on $A$.
On the other hand, if $U(\bm{\theta})$ transforms $X\otimes I^{\otimes n_A}$ to a nonlocal Pauli operator acting both on $A$ and $B$, $x_0$ vanishes in $\rho_0^A$ since the partial trace removes such a term.
If $U(\bm{\theta})$ acts in an intermediate manner of the above two extreme cases, the magnitude of $x_0$ in $\rho_0^A$ becomes smaller than the former case, that is, QRNN partially ``forgets'' $x_0$.
We believe that we can employ these insights to construct more efficient QRNNs in future research.

\subsection{Experimental realization}
As mentioned above, expectation values of an observable $\langle O \rangle$ are used as a new input in the prediction phase of QRNN, which requires us to obtain $\braket{O}$ in a single-shot manner.
Ensemble quantum systems such as NMR quantum computers, where such measurements are possible, are desirable for this reason.
While nuclear spins used as qubits in NMR are hard to initialize, we believe this problem can be resolved by e.g. a technique called dynamic nuclear polarization, which initializes nuclear spins by transferring the state of initialized electron spins \cite{deBoer1974, HENSTRA19906, Tateishi2014, PhysRevB.87.125207}.

On the other hand, the use of single quantum systems such as superconducting or ion trap qubits only allows us to extract bitstrings with single-shot measurements, which leads to inefficiency in the prediction phase for time-series prediction tasks.
It requires us to run the whole QRNN to some time step $t$ to obtain a prediction $\overline{\bm{y}_{t}}$, which must be obtained with some accuracy to proceed to the next time step since it is used as the input to the system for such tasks.
However, we might be able to extend the framework of QRNN to use such single quantum systems; for example, they can be used for time-series prediction of binary data, where the predicted data itself is represented by bitstrings.
We leave such an extension for future work.

\subsection{Relation between existing algorithms}
The quantum reservoir computing (QRC) \cite{Fujii2017} has been proposed as a technique to tackle temporal learning tasks using complex quantum dynamics.
QRC is a quantum-classical hybrid algorithm similar to QRNN from the viewpoint of the purpose; both of them are used to train and predict time-series data.
The difference between QRNN and QRC is that while QRC updates readout transformation $g$ to minimize the cost function, QRNN mainly tunes the quantum system by optimizing the circuit parameter $\bm{\theta}$.
Subsequent theoretical \cite{Chen2019, Chen2020, Ghosh2019, Ghosh2020, ghosh2020universal,govia2020quantum,Martinez2020,Kutvonen2020} and experimental works \cite{negoro2018machine, Chen2020} have extended the QRC in various ways.

We also notice a similar title in Ref. \cite{bausch2020recurrent}.
This approach is based on quantum neurons presented in Ref. \cite{cao2017quantum}, and use sophisticated quantum circuits.
This is in contrast to our approach, where the parametrized circuit $U(\bm{\theta})$ can be constructed in a hardware-efficient way, as we show in the next section by numerical simulations.

\section{Numerical experiments} \label{sec:numerical}

\begin{figure}
    \centering
    \includegraphics[width=0.6\linewidth]{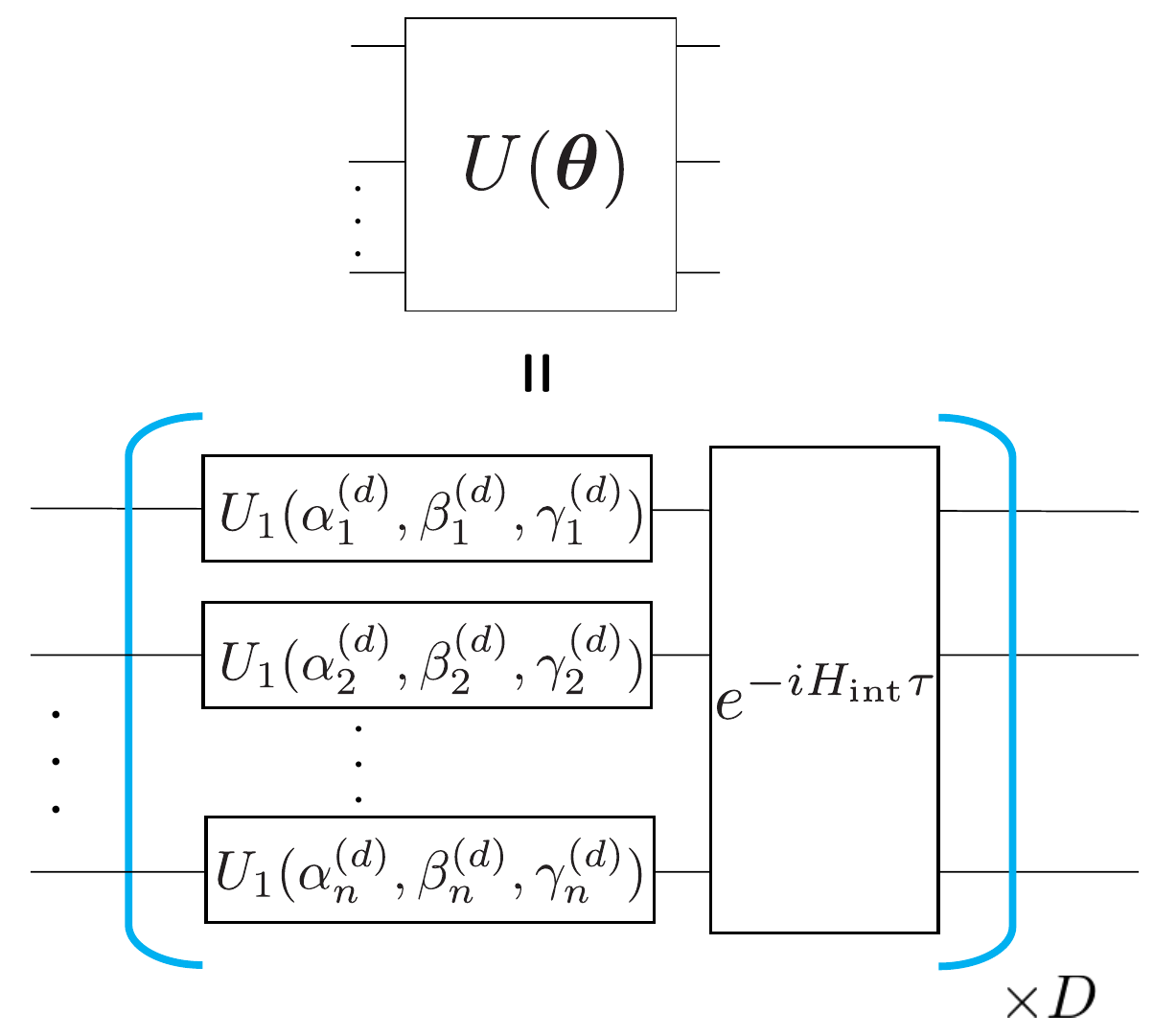}
    \caption{The evolution part of the circuit utilized in the numerical simulations.
    $\alpha_i^{(d)}$, $\beta_i^{(d)}$, and $\gamma_i^{(d)}$ are rotation angles and act as independent parameters of the circuit, i.e., $\bm{\theta}$ consists of $\{\alpha_i^{(d)}, \beta_i^{(d)}, \gamma_i^{(d)}\}$. $d=1,\cdots, D$ is the layer index.
    }
    \label{fig:rotation_gate}
\end{figure}

Here, we demonstrate and analyze the performance of the proposed QRNN by simulating its training and prediction phases for time-series prediction tasks.
We use scalar input data $\{x_t\}$, and the task is to construct a QRNN that can output $\overline{y_t}\approx x_{t+1}$.
Throughout the simulations presented in this section, we investigate a QRNN circuit using $n=6$ qubits; groups A and B have $n_A=3$ and $n_B=3$ qubits, respectively. 
As the encoding gate, we use,
\begin{align}
    U_{\rm in}(x_t)=R_y(\arccos{x_t}),
\end{align}
acting on each qubit in group B, following the ideas presented in Ref. \cite{Mitarai2018}.
Note that the input gate can be chosen differently; for example, we might be able to improve the performance of QRNN by changing it to encode orthogonal polynomials.
At the evolution part, the circuit shown in Fig. \ref{fig:rotation_gate} is applied as the parametrized gate $U(\bm{\theta})$, which consists of alternating layers of single-qubit rotations and Hamiltonian dynamics.
We fix the Hamiltonian and optimize the angles of the single-qubit rotations.
The single-qubit rotations are parametrized as, 
\begin{align}
    U_1(\alpha, \beta, \gamma)=R_x(\alpha)R_z(\beta)R_x(\gamma),
    \label{eq:X-Zdecomposition}
\end{align}
where $\alpha$, $\beta$, and $\gamma$ are real numbers, and $R_x$ and $R_z$ are single-qubit rotations around $x$ and $z$-axis, respectively.
After the rotation, the whole system is evolved with the following Hamiltonian,
\begin{align}
    H_{\rm{int}}=\sum_{j=1}^{n} a_{j} X_{j}+\sum_{j=1}^{n} \sum_{k=1}^{j-1} J_{j k} Z_{j} Z_{k},
    \label{eq:H_int}
\end{align}
following Ref. \cite{Fujii2017, Mitarai2018}.
We denote the evolution time by $\tau$.
We repeat applications of $U_1(\alpha, \beta, \gamma)$ and $e^{-iH_{\rm int} \tau}$ for $D=3$ times, as shown in Fig. \ref{fig:rotation_gate}.
The coefficients $a_j, J_{jk}$ are taken randomly from a uniform distribution on $[-1, 1]$ and fixed during the training.
At the measurement part, we measure $Z$-expectation value of each qubit in group B and take $\overline{y_t}$ as their average multiplied by a real coefficient $c$.
In the simulation presented in the following subsections, the coefficient $c$ is also optimized along with $(\alpha_i^{(d)}, \beta_i^{(d)}, \gamma_i^{(d)})$ for each qubit.
For classical optimizer, we employ the BFGS algorithm implemented in SciPy \cite{Virtanen2020}.
All of the parameters are initialized to $0$ except for the coefficient $c$ which is initialized to $1$.

\subsection{Demonstration}

First, we train the above described QRNN to predict cosine wave and triangular wave.
For the cosine wave, we use $x_t = \cos{(\pi t')}/2$, and for the triangular wave, we use
\begin{align}
    x_t = \begin{cases}
    -t'+\frac{1}{2} & (0 \leq t' \leq 1) \\
    t'-\frac{3}{2} & (1 \leq t' \leq 2) \\
    -t'+\frac{5}{2} & (2 \leq t' \leq 3) \\
    t'-\frac{7}{2} & (3 \leq t' \leq 4)
  \end{cases},
\end{align}
for $t'=\frac{8}{199}t$.
For both of them, $\{x_t\}_{t=0}^{99}$ is used as training data.
After the training, we test the accuracy of the predictions for $100\leq t < 200$.
The evolution time $\tau$ is set to $0.2$ in this numerical experiment.

Secondly, we train QRNN to predict quantum spin dynamics since we envision that the proposed approach is suited for such applications, that is, for predictions of quantum phenomena.
Here, we use open quantum dynamics of a 3-spin system generated by a Lindblad master equation,
\begin{align}
    \frac{d}{dt'}\sigma(t') &= -i[H, \sigma(t')] \notag \\
    &+ \sum_{k} \frac{1}{2}\left[2 C_{k} \sigma(t') C_{k}^{+}-\sigma(t') C_{k}^{\dagger} C_{k}-C_{k}^{\dagger} C_{k} \sigma(t')\right]
\end{align}
with $H = -\frac{1}{2}\sum^{3}_{i=1}h^{(i)} Z_i -\frac{1}{2}\sum^{2}_{i=1}(J_x^{(i)}X_i X_{i+1} + J_y^{(i)}Y_i Y_{i+1}+J_z^{(i)}Z_i Z_{i+1})$ and $\{C_k\} = \{c(X_i+Y_i)\}_{i=1}^3$ as the data to be predicted.
The coefficients are set to $h^{(i)}=2\pi$, $J_\mu^{(i)} = 0.1\pi$ for $\mu=x, y, z$, and $c=\sqrt{0.002}$.
The initial state of the system is chosen to be $\ket{0}^{\otimes 3}$.
We train the QRNN to predict the $X$-expectation value of the first spin, $\braket{X_1(t')}$, at time $t'=\frac{100}{499}t$, that is, we take $x_t=\braket{X_1(t')}$.
$\{x_t\}_{t=0}^{199}$ is used for the training, and we test the prediction afterwards for $200\leq t<500$. The evolution time $\tau$ is set to $0.18$ in this numerical experiment.

\begin{figure}
    \centering
    \includegraphics[width=\linewidth]{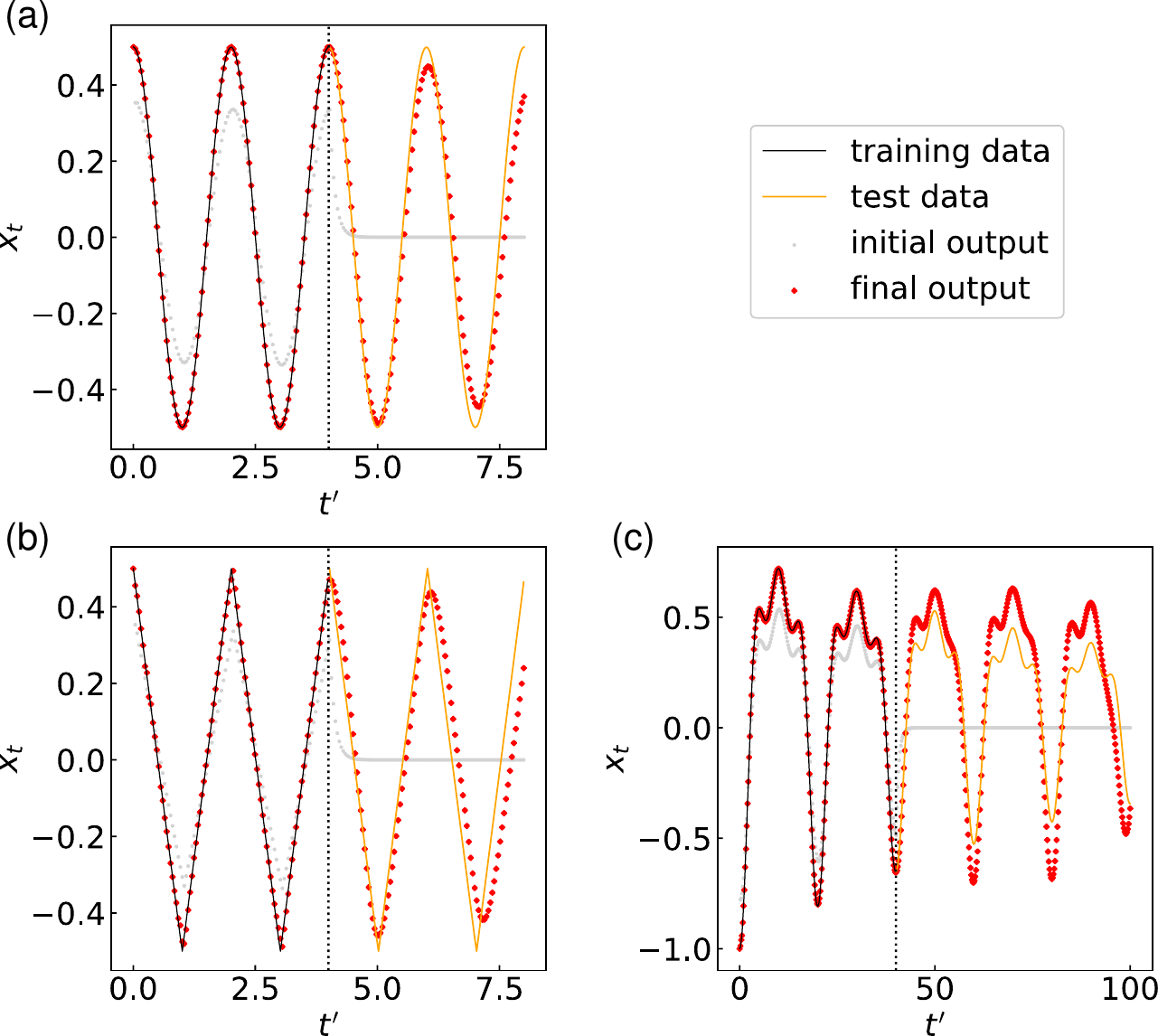}
    \caption{Results of numerical simulation of training the QRNN to predict (a) cosine wave, (b) triangular wave, and (c) three-spin Lindblad dynamics.
    The data points on the left of the vertical grey dashed line are used as the training dataset.
    The solid lines represent the function which the QRNN is trained to model.
    The parts of the data used in the training phase and the prediction phase are depicted as the black and orange lines, respectively.}
    \label{fig:results}
\end{figure}

Figure \ref{fig:results} shows the results of the numerical simulations whose mean squared error (MSE) between the output and the teacher on the first 25 test points are the smallest among 10 randomly generated coefficients $a_{j}, J_{jk}$.
In the figure, the initial output shows a sequence of predictions $\{\overline{y_t}\}$ obtained from QRNN with the initial parameters.
The reason it resembles the true data $x_t$ in the training region is because we input $x_t$ by $R_y(\arccos{x_t})$, which encodes $x_t$ to $Z$-expectation values of qubits.
This means that $x_t=\overline{y_{t}}$ if $\tau=0$.
For nonzero $\tau$, we can naturally expect $\overline{y_{t}}$ to be somewhat smaller than $x_t$ because of the interaction.
Looking at the optimized output, we can see that QRNN can be successfully trained to fit a given training dataset and to make predictions after the training.
This can also be verified with MSE between the output and the teacher evaluated on the first 25 test points, which is on the order of $10^{-3}$ for all data types (Tab. \ref{tab:mse}).
During the experiments, we found that the quality of the prediction strongly depends on the parameter $\tau$ which determines the interaction strength between the qubits.
This phenomenon is further analyzed in the next subsection.

\begin{table}
    \caption{\label{tab:mse} Mean squared error between the trained output and the test data shown in Fig. \ref{fig:results} evaluated with the first 25 points after the training region.}
    \begin{ruledtabular}
        \begin{tabular}{cccc}
        Data & cos & triangle & spin dynamics\\ 
        MSE & $3.33\times 10^{-4}$ & $2.6\times 10^{-3} $ & $2.95\times 10^{-3}$
        \end{tabular}
    \end{ruledtabular}
\end{table}

\subsection{Dependence on the interaction strength}
Here, we vary $\tau$ from 0 to 10 and perform the training of the QRNN on the cosine wave used in Fig. \ref{fig:results} (a) to investigate the dependency of the performance of the QRNN on the interaction strength.
For this purpose, we randomly drew the coefficients of the interaction Hamiltonian $H_{\rm{int}}$ (Eq. (\ref{eq:H_int})) 10 times from uniform distribution on $[-1, 1]$, and trained each QRNN.

\begin{figure}
    \centering
    \includegraphics[width=\linewidth]{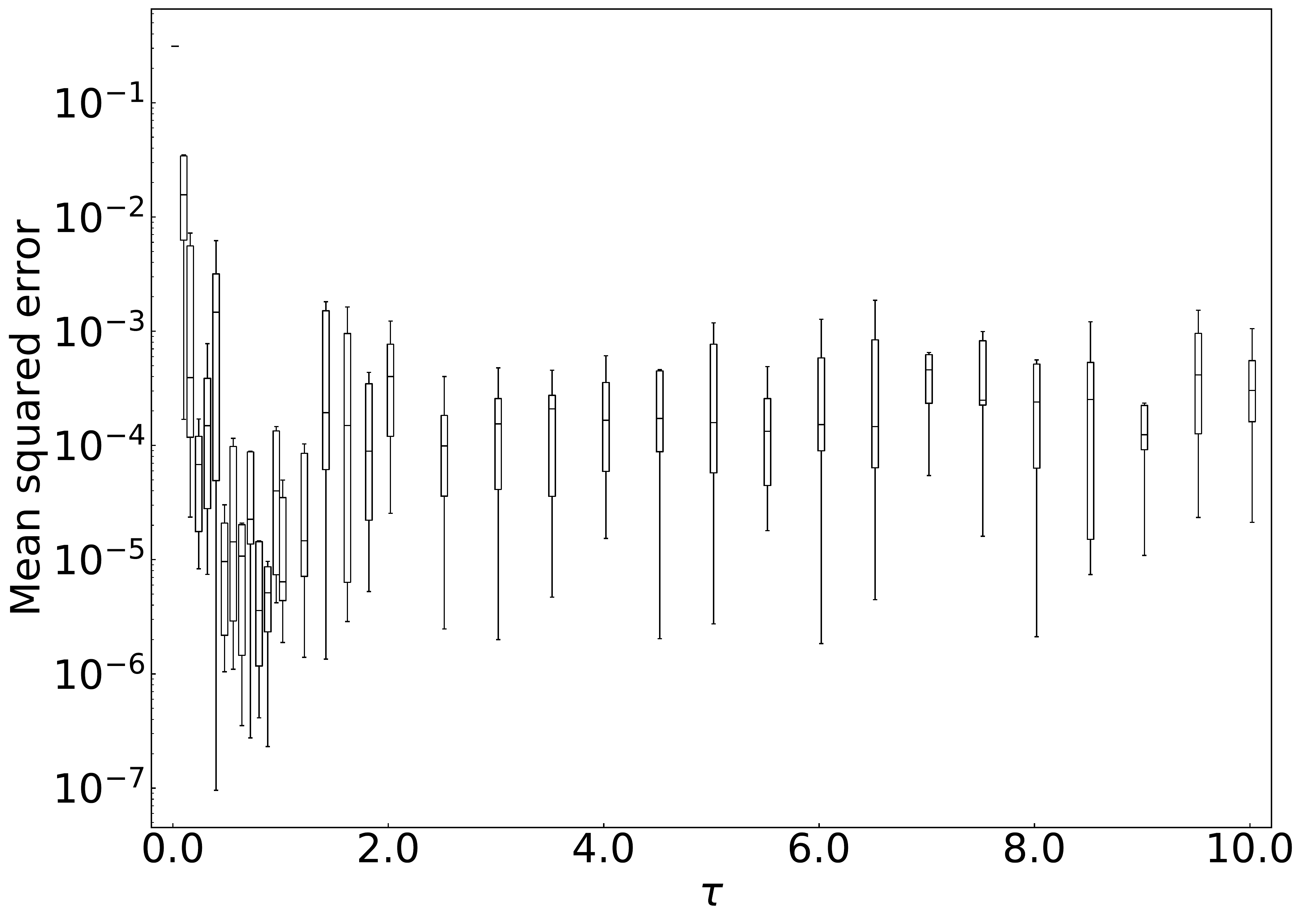}
    \caption{Relation between the evolution time $\tau$ and the prediction error evaluated in terms of the mean squared error.}
    \label{fig:tau_vs_mse}
\end{figure}

Fig. \ref{fig:tau_vs_mse} shows the mean squared error between $x_t$ and $\overline{y_t}$ of the first 25 time steps.
From Fig. \ref{fig:tau_vs_mse}, we notice that there exists a certain range of $\tau$ which provides us relatively accurate predictions.
More specifically, the QRNN circuits with $\tau\approx a_j,~J_{ij}$ work better than the other ones.
This is as expected because a short $\tau$ leads to less information transferred to the subsystem $A$.
It means that the QRNN cannot store the past information, since the subsystem $B$ is measured and initialized at every time step, which erases all information that is left in $B$.
On the other hand, when $\tau$ is sufficiently larger than $a_j$ and $J_{ij}$ of the Hamiltonian, the dynamics under $H_{\mathrm{int}}$ becomes complex, which makes it hard for the single-qubit rotations to extract the required information from the system.
This analysis shows that the QRNN indeed utilizes the past information stored in the subsystem $B$, and we must have a certain amount of entangling gates at the evolution part of the QRNN.

\section{Conclusion} \label{sec:conclusion}

We proposed a quantum version of an RNN for temporal learning tasks.
The proposed algorithm, QRNN, employs a parametrized quantum circuit with a recurrent structure.
In QRNN, one group of qubits are measured and initialized at every time step to obtain outputs and inject inputs, while other qubits are never measured to store the past information.
We provided numerical experiments to verify the validity of the idea.

It remains an open question whether the proposed QRNN performs better than the classical RNN, and we leave it as a future research direction.
However, we note that the QRC which has a similar structure as the QRNN has been shown to have better performance than its classical counterpart \cite{Fujii2017}.
We can, therefore, expect the QRNN to have similar performance when its structure is optimized.
As the QRNN explicitly optimize the dynamics of the quantum system, we especially expect it to be able to predict quantum phenomena, such as the one we used in our numerical experiments.

\vspace{5mm}

\begin{acknowledgements}
KM is supported by JSPS KAKENHI No. 19J10978, 20K22330 and JST PRESTO JPMJPR2019.
KF is supported by KAKENHI No.16H02211, JST PRESTO JPMJPR1668, JST ERATO JPMJER1601, and JST CREST JPMJCR1673.
MN is supported by JST PRESTO JPMJPR1666.
This work is supported by MEXT Quantum Leap Flagship Program (MEXT Q-LEAP) Grant Number JPMXS0118067394 and JPMXS0120319794.
\end{acknowledgements}

\appendix*

\section{Calculation of analytic gradient} \label{apdx:derivative}
Here we describe how to evaluate analytic gradient in the QRNN using the so-called parameter shift rule \cite{Mitarai2018, schuld2019evaluating, koczor2020quantum, Mitarai2019meth}.
In short, the rule allows us to evaluate an expectation value of any observable $O$ with respect to the following operator,
\begin{align}
\frac{\partial}{\partial \theta_i}\left[U(\bm{\theta})\rho U^\dagger(\bm{\theta})\right] = \frac{\partial U(\bm{\theta})}{\partial \theta_i}\rho U^\dagger(\bm{\theta}) + U(\bm{\theta})\rho \frac{\partial U^\dagger(\bm{\theta})}{\partial \theta_i},
\end{align}
for any state $\rho$ by evaluating the expectation value twice at shifted parameters if the parametrized gate corresponding to the parameter $\theta_i$ can be written as $e^{i\theta_i P}$ for some Pauli operator $P$.

For simplicity, we assume $x_t$ to be scalar and that the output is obtained from a single observable $O$ as $\overline{y_t}=g\left(\braket{O}_t\right)$.
Also, the cost function is assumed to be  $L=\frac{1}{2}\sum_{t=0}^{T-2}(\overline{{y}_{t}}-{x}_{t+1})^{2}$. However, generalization of the discussion below to other cost functions is straight forward.
Now, the gradient of $L$ can be written as,
\begin{align}\label{appeq:dl}
    \frac{\partial L}{\partial \theta_i} &= \sum_{t=0}^{T-2}(\overline{y_t}({\bm \theta}, x_0, \cdots x_{t})-x_{t+1})\frac{\partial \overline{y_{t}}}{\partial \theta_i}.
\end{align}
Next, we can express $\frac{\partial \overline{y_t}}{\partial \theta_i}$ as,
\begin{align}\label{appeq:dx}
    \frac{\partial \overline{y_t}}{\partial \theta_i} &= \frac{\partial g\left(\braket{O}_{t}\right)}{\partial \theta_i} = \frac{\partial g}{\partial \braket{O}_{t}}\frac{\partial \braket{O}_{t}}{\partial \theta_i}.
\end{align}
Let us define the input state $\rho_{\mathrm{in}, t}^{B} := U_{\mathrm{in}}(x_{t})\ket{0}\bra{0}U^\dagger_{\mathrm{in}}(x_{t})$ and $\rho_{\mathrm{in},t}^{AB}(\bm{\theta},x_0,\cdots,x_{t}) = \rho_{t-1}^A\otimes \rho_{\mathrm{in}, t}^{B}$.
Below we abbreviate the dependence of $\rho_{\mathrm{in},t}^{AB}$ on $x_0,\cdots,x_{t}$ and just write $\rho_{\mathrm{in},t}^{AB}(\bm{\theta})$ for simplicity.
Then, $\frac{\partial \braket{O}_{t}}{\partial \theta_i}$ can be written as, 
\begin{align}
    \frac{\partial \braket{O}_{t}}{\partial \theta_i} &=  \frac{\partial}{\partial \theta_i}\Tr \left[\rho^B_t O\right] \notag \\
    &= \frac{\partial}{\partial \theta_i} \Tr \left[ U (\bm{ \theta}) \rho_{\mathrm{in},t}^{AB}(\bm{\theta})  U^\dagger (\bm{\theta}) O\right] \notag \\
    \begin{split}
        &= {\rm Tr}\left[\frac{\partial U({\bm \theta})}{\partial \theta_i} \rho_{\mathrm{in},t}^{AB}({\bm \theta}) U^{\dagger}({\bm \theta}) O\right]\\
        &\quad+ {\rm Tr}\left[U({\bm \theta}) \rho_{\mathrm{in},t}^{AB}({\bm \theta}) \frac{\partial U^{\dagger}({\bm \theta})}{\partial \theta_i} O\right] \\
        &\quad+ {\rm Tr} \left[U({\bm \theta}) \frac{\partial \rho_{\mathrm{in},t}^{AB}(\bm{\theta})}{\partial \theta_i} U^{\dagger}({\bm \theta})O\right]. \label{appeq:do}
    \end{split}
\end{align}
In Eq. (\ref{appeq:do}), the first two terms can easily be evaluated by the parameter shift rule.
As for the last term, 
\begin{align}
    \frac{\partial \rho_{\mathrm{in},t}^{AB}({\bm \theta})}{\partial \theta_i} &=  \frac{\partial \rho_{t-1}^A({\bm \theta})}{\partial \theta_i} \otimes \rho_{\mathrm{in},t}^B \notag \\
    &= \frac{\partial}{\partial \theta_i}\Tr_B \left[U({\bm \theta})\rho_{\mathrm{in},t-1}^{AB}(\bm{\theta}) U^{\dagger}({\bm \theta})\right] \otimes \rho_{t}^B \notag \\
    \begin{split}
    &= \left\{\Tr_B \left[ \frac{\partial U({\bm \theta})}{\partial \theta_i} \rho_{\mathrm{in},t-1}^{AB}(\bm{\theta}) U^{\dagger}({\bm \theta})\right]\right. \\
    &\quad+ \Tr_B\left[U({\bm \theta}) \rho_{\mathrm{in},t-1}^{AB} (\bm{\theta})\frac{\partial U^{\dagger}({\bm \theta})}{\partial \theta_i}\right] \\
    &\quad \left.+ \Tr_B\left[U({\bm \theta}) \frac{\partial \rho_{\mathrm{in},t-1}^{AB}(\bm{\theta})}{\partial \theta_i} U^{\dagger}({\bm \theta})\right]\right\} \otimes \rho_{t}^B.     
    \end{split}
\end{align}
In the above, the first two terms can be evaluated with parameter shift rule.
For the last term, we need to expand it again to express $U({\bm \theta}) \frac{\partial \rho_{\mathrm{in},t-1}^{AB}(\bm{\theta})}{\partial \theta_i} U^{\dagger}({\bm \theta})$ in terms of $U({\bm \theta}) \frac{\partial \rho_{\mathrm{in},t-2}^{AB}(\bm{\theta})}{\partial \theta_i} U^{\dagger}({\bm \theta})$.
The above discussion implies that, to evaluate $U({\bm \theta}) \frac{\partial \rho_{t}^{AB}({\bm \theta})}{\partial \theta_i} U^{\dagger}({\bm \theta})$ which appears in Eq. (\ref{appeq:do}), we have to perform the above expansion reccursively down to $U({\bm \theta}) \frac{\partial \rho_{\mathrm{in},0}^{AB}}{\partial \theta_i} U^{\dagger}({\bm \theta})$, where the expansion terminates since $\rho_{\mathrm{in},0}^{AB}=\ket{0}\bra{0}\otimes \rho^{B}_{\mathrm{in}}(x_0)$ does not depend on $\bm{\theta}$.
As every recursive expansion yields a pair of terms that can be evaluated by shifting the parameter twice, we need $2T$ additional evaluations of $\braket{O}$ to obtain the analytical gradient of the cost function for each $\theta_i$.

\end{document}